\documentstyle[12pt]{article}
\begin{document}
\title{ON THE MEANING OF 	 GENERAL COVARIANCE  AND THE RELEVANCE OF  OBSERVERS  IN GENERAL RELATIVITY }
\author{L. Herrera$^1$\thanks{On leave from UCV, Caracas, Venezuela, e-mail: lherrera@usal.es}\\
{$^1$Departamento de  de F\'{\i}sica Teorica e Historia de la  F\'{\i}sica,}\\
Universidad del Pais Vasco, Bilbao, Spain}
\maketitle
Essay written for the Gravity Research Foundation 2011 award Essays on Gravitation and receiving  Honorable Mention.\\
\begin{abstract}
Since the appearance of General Relativity, its intrinsec general covariance has been very often  misinterpreted as implying that physically meaningful quantitities (and conclusions extracted from the theory) have to be absolutely independent on observers. This incorrect point of view is sometimes expressed by discarding the very concept of observer in the structure and applications of the theory. As we shall stress in this essay, through some examples, the concept of observer is as essential to General Relativity as it is  to any physical theory.  
\end{abstract}

{\it Entia non sunt multiplicanda praeter neccessitatem}
(Entities should not be introduced except when strictly necessary).
William van Ockham.

\section{INTRODUCTION}
Observers play an essential role in any physical theory. This is particularly true in Quantum Mechanics where the very concept of reality is tightly attached to the existence of the observer, as ingeniously expressed by the well known riddle "If a tree falls in a forest and no one is around to hear it, does it make a sound? 

However, with the emergence of General Relativity and its instrinsec general covariance, the idea has arisen in some people's mind, according to which general covariance is understood as the statement that {\it all observers are physically equivalent}. This misinterpretation too often supersedes the right one, namely {\it when written down covariantly, the equations describing any physical theory may be used by any observer}. Thus for example if one writes the Newtonian equation of motion in terms of ``Galilean'' three vectors then rotating observers using that equation will obtain absurd results  unless some fictitious ``inertial forces'' are introduced ad hoc. However Newtonian equation of motion can be written in a generally covariant form by introducing affine conections ( see Trautman lectures in \cite{Brandeis}) and in that form it can be used by any (rotating included) observer without the introduction of artifacts such as ``inertial forces''. To summarize this point: general covariance of a given theory does not mean that all observers are physically equivalent, it only means that  any observer is entitled to  use it. 

The first consequence of the confusion described above is that the very concept of observer becomes deprived of any physical meaning. Indeed, if one assumes that all observers are physically equivalent, this concept loses all its relevance when defining fundamental quantities and describing a specific physical scenario. 

However, in the structure of any physical theory, the variables in terms of which those theories are expressed are usually defined for a given congruence of 
 observers. Thus for example in hydrodynamics when we speak about pressure, energy density, entropy, temperature, etc of any fluid element, we refer to the value of those quantities as measured by a specific observer. The reason of this being that the measured value of such quantities depend on the  velocity of the observer with respect to the fluid element and accordingly it is mandatory to specify the observer, the most obvious choice being (in this particular example) the comoving observer.

A similar  situation is found in Mawxell electrodynamics, which usually is described in terms of the two tree--vectors {\bf E} and {\bf B}, from which as is well known a four--vector cannot be defined. However introducing a congruence of preferred observers whose world lines are defined by the four vector ${\bf u}$ tangent to those world  lines, two  four--vectors  can be constructed as 
\begin{equation}
E^{\alpha}=F^{\alpha \beta}u_{\beta}, \qquad B_{\alpha}=F^{*}_{\alpha \beta}u^{\beta},
\label{1}
\end{equation}
where $F^{\alpha \beta}$ is the Maxwell tensor and $F^{*}_{\alpha \beta}$ its dual.
In a preferred cartesian frame the components of the above four--vectors coincide with {\bf E} and {\bf B} respectively.
From this example it appears that  these newly defined quantities are invariants (tensors) at the price of introducing  an ``auxiliary'' vector field, which could  be interpreted as a disadvantage. However this is not so, indeed we agree with Trautman \cite{Brandeis} in that one should not introduce additional structures besides those alredady present in the axioms of the theory and those that are necessary to describe the physical systems. Our point is that   vector field ${\bf u}$ belongs to that category of structures, and its introduction vindicates the role of the observer  in the theory  thereby enhancing its richness. 

It should be mentioned that the main  line of arguments presented above is not new and  has been sustained  long before by many important relativists such as Fock, M\o ller and Bondi among others. We shall endorse here that point of view, reinforcing it by means of some examples and  presenting    some invariant quantities (defined in terms of tensors) which are defined with respect to specific congruence of observers and which play an important role in the description of self--gravitating objects, bringing out the relevance of the formers.
\section{Superenergy}
On of the most important concepts in general relativity involving the  congruence of observers is  superenergy (and related quantities) \cite{bel1}--\cite{bel4}. 

As  is known, in classical field theory, energy is a quantity
defined in terms of potentials and their first derivatives. In
General Relativity however, it is impossible to construct a tensor
expressed only through the metric and their first derivatives (due to the
equivalence principle). Accordingly, a local description of
gravitational energy in terms of true invariants (tensors of any
rank) is  not possible within the context of the theory.

Thus, one is left with the following three alternatives:
\begin{itemize}
\item  Looking for a non--local  definition of energy.
\item  Finding a definition based on  pseudo--tensors.
\item  Resorting to a succedaneous definition, e.g.:  superenergy.
\end{itemize}

One example of the last of the above alternatives is superenergy,   which may be defined
from the Bel  or the  Bel--Robinson tensor 
(they both coincide in vacuum), and has been shown to be very useful
when it comes to explaining  a number of  phenomena in the context of
general relativity.

Thus, for instance, it helps to explain the occurrence of vorticity
in  both  radiative  \cite{HB}, and stationary spacetimes \cite{HC}.
Also, it renders intelligible the behaviour of test particles moving
in circles around the symmetry axis in an Einstein--Rosen spacetime
\cite{HE}.

Both the Bel and the Bel--Robinson tensors are obtained from the Riemann and the Weyl tensor (as well as their dual) respectively  by analogy  with  the form on which the energy--momentum tensor of the electromagnetic field depends on the Maxwell tensor (and its dual).

An important role in the theory underlying  the concept of superenergy is played by  three observer dependent  tensors emerging from the orthogonal splitting of the Riemann tensor.

Thus following Bel\cite{bel3},(see also  \cite{parrado} for more recent references), let us introduce the following tensors:
\begin{equation}
Y_{\alpha \beta}=R_{\alpha \gamma \beta \delta}u^{\gamma}u^{\delta},
\label{electric}
\end{equation}
\begin{equation}
Z_{\alpha \beta}=^{*}R_{\alpha \gamma \beta
\delta}u^{\gamma}u^{\delta}= \frac{1}{2}\eta_{\alpha \gamma
\epsilon \rho} R^{\epsilon \rho}_{\quad \beta \delta} u^{\gamma}
u^{\delta}, \label{magnetic}
\end{equation}
\begin{equation}
X_{\alpha \beta}=^{*}R^{*}_{\alpha \gamma \beta \delta}u^{\gamma}u^{\delta}=
\frac{1}{2}\eta_{\alpha \gamma}^{\quad \epsilon \rho} R^{*}_{\epsilon
\rho \beta \delta} u^{\gamma}
u^{\delta},
\label{magneticbis}
\end{equation}
with
$R^{*}_{\alpha \beta \gamma \delta}=\frac{1}{2}\eta_{\epsilon \rho \gamma \delta}R_{\alpha \beta}^{\quad \epsilon \rho}$ and $\bf u$ denoting the timelike vector field associated to  the congruence of observers. The first two tensors define  the ``electric'' and ``magnetic'' part of the Riemann tensor, the third one has no analogy in electrodynamics.

It can be shown that the Riemann tensor  can be expressed through  these tensors in what is called the orthogonal splitting of the Riemann tensor (see \cite{parrado} for details).

 As we shall see next a family of scalar  functions (hereafter referred to as structure scalars) may be defined from the three tensors mentioned above. The physical relevance of such scalars, which explicitely depend on a given congruence of observers, support further the case for the physical relevance of the latter.

\section{Structure scalars}
We shall now introduce five scalars quantities \cite{H1} explicitly involving the congruence of observers and whose physical meaning will strength  further the case for observer dependent variables.

Let us first notice  that  tensors  $X_{\alpha \beta}$ and $Y_{\alpha \beta}$ can be splitted in terms of their traces and  the corresponding trace--free tensor, i.e.
\begin{equation}
X_{\alpha \beta}=\frac{1}{3}Tr X h_{\alpha \beta}+ X_{<\alpha \beta>},
\label{esn}
\end{equation}
with $Tr X=X^\alpha_\alpha$ and,
\begin{equation}
X_{<\alpha \beta>}=h^\mu_\alpha h^\nu_\beta(X_{\mu \nu}-\frac{1}{3}Tr X h_{\mu \nu}),
\label{esnII}
\end{equation}
where $h^\mu_\alpha$ is the projector on the hypersurface orthogonal to {\bf u}.

In the case of  spherically symmetric distributions of collapsing
fluid,  assumed to be locally anisotropic, and 
undergoing dissipation in the form of heat flow (diffusion limit)  and/or free streaming  radiation  it can be shown that \cite{H1}

\begin{equation}
TrX\equiv X_T=8\pi ( \rho+\epsilon),
\label{esnIII}
\end{equation}
and
\begin{equation}
X_{<\alpha \beta>}=X_{TF}(s_\alpha s_\beta+\frac{h_{\alpha \beta}}{3}),
\label{esnIV}
\end{equation}
where  
\begin{equation}
X_{TF}\equiv (4\pi \Pi-E),
\label{esnIVn}
\end{equation}
with $\rho$ and  $\epsilon$ denoting the energy--density of the fluid  and the energy--density of the null fluid describing the dissipation in the streaming out limit, respectively. Also, $E$ denotes a scalar in terms of which the ``electric'' part of the Weyl tensor may be defined (the ``magnetic'' part of the Weyl tensor vanishes due to spherical symmetry), $s^\alpha$ is a  unit spacelike vector field orthogonal to {\bf u} and $\Pi$ denotes the local anisotropy of pressure.

In a similar way it can be obtained
\begin{equation}
TrY\equiv Y_T=4\pi[(\rho+\epsilon)+3( P_r+\epsilon)-2\Pi],
\label{esnV}
\end{equation}
and
\begin{equation}
Y_{<\alpha \beta>}=Y_{TF}(s_\alpha s_\beta+\frac{h_{\alpha \beta}}{3}),
\label{esnVI}
\end{equation}
with  
\begin{equation}
Y_{TF}\equiv (4\pi\Pi+E).
\label{defYTF}
\end{equation}
Finally a fifth scalar may be defined   as 
\begin{equation}
Z=
\sqrt{Z_{\alpha\beta}Z^{\alpha\beta}}=\frac{8\pi}{\sqrt{2}}(q+\epsilon),
\label{zz}
\end{equation}
where $q$ is a scalar function in terms of which the heat flow vector (describing dissipation in the diffusion approximation) can be expressed.

From the above it follows that local anisotropy of pressure is  determined by  $X_{TF}$ and $Y_{TF}$
by
\begin{equation}
8\pi \Pi=X_{TF} + Y_{TF}.
\label{defanisxy}
\end{equation}

Some of the more relevant  physical properties of the above scalars are:
\begin{itemize}
\item The physical meaning of  $X_T$  and $Z$ is evident and does not require further clarification. 
\item  In the absence of dissipation,   $X_{TF}$ controls inhomogeneities in the energy density \cite{H1}. 
\item $Y_{TF}$ describes the influence of the local anisotropy of pressure and density inhomogeneity on the Tolman mass \cite{H1}
\item $Y_T$ appears to be proportional to the  Tolman mass ``density'' for systems in equilibrium or quasi--equilibrium \cite{H1}.
\item The evolution of the expansion scalar and the shear tensor  is fully controlled by $Y_{TF}$ and $Y_T$ \cite{H1, H2, H3}
\end{itemize}

We shall next discuss the role of these scalars in the interpretation of a given solution to Einstein equations.

\section{Exact solutions and their observer depending interpretation}
It is already a stablished fact that  a variety of line elements may satisfy the Einstein equations for different (physically meaningful) stress--energy tensors (see\cite{1}--\cite{5} and references therein). 

Particularly interesting is the situation when the two possible interpretations of  a given spacetime correspond to a boost of  one of the observer congruence with respect to the other. Obviously the structure scalars for both congruences  of observers would be different, implying that the physical properties described by such scalars would differ too

This is for example the case of the zero curvature FRW model, which represents a perfect fluid solution for observers at rest with respect to the timelike congruence defined by the eigenvectors of the Ricci tensor, but can also be interpreted as the exact solution for a viscous dissipative fluid as seen by  observers moving relative to the previously mentioned congruence of observers \cite{2}. An important point to mention  is that the relative (``tilting'') velocity between the two congruences may be related to a physical phenomenon such as the observed motion of our galay relative to the microwave background radiation.

 From these comments  and the physical meaning of structure scalars enumerated in the previous section it follows  that zero curvature FRW models as described by ``tilted'' observers will detect dissipative processes, energy--density inhomogeneity, different evolution of the expansion scalar and the shear tensor, among other differences, with respect to the ``standard'' observer (see \cite{2} for a comprehensive discussion on this example). 

Another example which is particularly enlightening, on the issue under consideration, is that of tilted Lemaitre--Tolman--Bondi (LTB)  spacetime. Indeed, due to the absence of conformal Killing vectors in a general LTB spacetime, it follows that the heat flow vector detected by the tilted observer is related to a real (non--reversible) dissipative process (see \cite{tilted} for details).

It should be emphasized that the key issue is not: what is the ``correct'' interpretation  of the model? since both are physically viable. The point is that  each  interpretation is related to a specific congruence of observers, and  the subjective element  ensuing from any specific choice  should not be taken as weakness of the theory but quite the opposite as  expression of its richness.

\section{Conclusions}
In this essay we wanted to emphasize three  points:
\begin{itemize}
\item  The equations of a covariantly defined theory are valid for all observers. This statement by no means implies that all observers are physically equivalent, in the same way as the fact that we  are all equal (in principle!) before the law does not imply that we all are equal (in general).
\item In all physical theories, part of the description of any specific scenario is related to the specific congruence of observers carrying out the study.  Such observer dependent element does not represent a drawback of the theory but  is a natural constituent in any physical  theory .
\item In the specific case of general relativity we have exhibited certain observer related quantities defined in terms of invariants and which play an important role in the description of self--gravitating objects. The advantage of such quantities is that they combine the information from the physical phenomenon with the information about the characteristics of the observer. Presented  examples (which by far are not exhaustive) reinforce further the two points above.
\end{itemize}


\begin{thebibliography}{99}
\bibitem{Brandeis} A. Trautman {\it Lectures on General Relativity--Brandeis Summer School} (Prentice Hall, New Jersey) (1964).
\bibitem{bel1} L. Bel {\it C. R. Acad. Sci. Paris} {\bf 247}, 1094 (1958).
\bibitem{bel2} L. Bel  {\it C. R. Acad. Sci. Paris} {\bf 248},  1297 (1959).
\bibitem{bel3} L. Bel  {\it Ann. Inst. H Poincar\'e}  {\bf 17}, 37 (1961).
\bibitem{bel4} L. Bel {\it Cah. de Phys.} {\bf 16}, 59 (1962);  {\it Gen. Rel. Grav.} {\bf 32}, 2047 (2000).
\bibitem{HB} L. Herrera, W. Barreto, J. Carot and   A. Di Prisco.
{\it Class.  Quantum Grav.}, {\bf 24}, 2645 (2007).

\bibitem{HC} L. Herrera, A Di Prisco and  J. Carot.
{\it Phys.  Rev. D} {\bf  76},  044012 (2007).

\bibitem{HE} L. Herrera, A Di Prisco, J. Carot and  N.O. Santos.
{\it Int. J.  Theor. Phys.} {\bf  47},  380 (2008).
\bibitem{parrado} A. Garc\'ia--Parrado Gomez Lobo, {\it arXiv:0707.1475v2}.
\bibitem{H1}  L. Herrera, J. Ospino, A. Di Prisco, E. Fuenmayor and O. Troconis, {\it Phys. Rev. D} {\bf 79}, 064025 (2009).
\bibitem{H2} L. Herrera, A. Di Prisco, J. Ospino and J. Carot {\it Phys. Rev. D} {\bf 82}, 024021 (2010).
\bibitem{H3} L. Herrera, A. Di Prisco and J. Ospino {\it Gen.Rel. Grav.} {\bf 42}, 1585 (2010).
\bibitem{1} A. R. King and G. F. R. Ellis {\it Commun. Math. Phys.} {\bf 31}, 209 (1973).
\bibitem{2} A. A. Coley and B. O. J. Tupper {\it Astrophys. J} {\bf 271}, 1 (1983).
\bibitem{3}  A. A. Coley and B. O. J. Tupper {\it Gen. Rel. Grav.} {\bf 15}, 977  (1983).
\bibitem{4}  A. A. Coley and B. O. J. Tupper {\it Phy. Lett. A} {\bf 100}, 495  (1984).
\bibitem{5} M. Calvao and J. M. Salim  {\it Class. Quantum Grav.} {\bf 9}, 127 (1992).
\bibitem{tilted} L. Herrera, A. Di Prisco and J. Iba\~nez {\it Phys. Rev. D} {\bf 84}, 064036 (2011).
\end{thebibliography}
\end{document}